\documentclass[twocolumn]{jpsj2}

\def\runtitle{Magnetic Structure of Nano-Graphite M\"{o}bius Ribbon}

\def\runauthor{Katsunori \textsc{Wakabayashi} and Kikuo \textsc{Harigaya}}

\def\mbf(#1){\mbox{\boldmath $#1$}}
\def\mob{M\"{o}bius }

\title{Magnetic Structure of Nano-Graphite M\"{o}bius Ribbon}

\author{
Katsunori \textsc{Wakabayashi}$^{1}$\thanks{Email address: 
waka@qp.hiroshima-u.ac.jp}, and
Kikuo \textsc{Harigaya}$^{2,3}$
}

\inst{%
$^1$Department of Quantum Matter Science,
Graduate School of Advanced Sciences of Matter (ADSM), 
Hiroshima University, Higashi-Hiroshima 739-8530, Japan \\
$^2$Nanotechnology Research Institute, AIST, Tsukuba 305-8568, Japan\\
$^3$Synthetic Nano-Function Materials Project, AIST, Tsukuba 305-8568,Japan
}

\recdate{\today}

\abst{
We consider the electronic and magnetic properties of nanographite
ribbon with zigzag edges under the periodic or M\"{o}bius boundary
conditions. The zigzag nano-graphite ribbons possess edge localized
states at the Fermi level which cause a ferrimagnetic spin polarization
localized at the edge sites even in the very weak Coulomb interaction. 
The imposition of the \mob boundary condition makes the system
non-AB-bipartite lattice, and depress the spin 
polarization, resulting in the formation of a magnetic domain wall.
The width of the magnetic domain depends on the Coulomb interaction
and narrows with increasing $U/t$. 
}

\kword{%
nanographite, graphite ribbon, \mob strip, Hubbard model
}

\begin{document}
\sloppy
\maketitle

Carbon based nano-scale materials such as fullerenes and carbon nanotubes
are attracting much attention due to their novel
electronic properties\cite{review1,review2}.
In these systems, the geometry of {\it sp}$^2$ carbon networks
crucially affects the electronic states near the Fermi level\cite{stm}.
The electronic states of carbon nanotubes are classified by 
the chiral vector which assigns the diameter and chirality of
the nanotubes.
Besides the closed carbon molecules, systems with open boundaries also
display unusual features connected with their shape.\cite{peculiar,waka}  
Nanographite ribbons, 
one-dimensional graphite lattices with a finite width,
have shown that ribbons with zigzag edges (zigzag ribbon) possess
localized edge states 
near the Fermi level \cite{peculiar,waka}. 
These edge states correspond to non-bonding molecular orbitals. Such
states are completely absent for ribbons with armchair edges.

While a graphite sheet behaves like a zero-gap semiconductor with vanishing 
the density of states (DOS) at the Fermi level, the edge states of the 
zigzag ribbons introduce a sharp
peak in the DOS at the same energy. Therefore, we expect that a 
strong Fermi instability might be induced by electron-phonon 
and/or electron-electron
interactions. The study of the electron-phonon interaction 
based on the Su-Schrieffer-Heeger (SSH) model concluded that
the lattice in-plane 
distortion does not occur in the zigzag ribbons,
because of the non-bonding character of edge state\cite{SSH}.
On the other hand, by treating the Hubbard model within the 
unrestricted Hartree-Fock (HF) mean-field approximation,  
the electron-electron interaction causes
the ferrimagnetic spin polarization at the zigzag edge
even under a very weak on-site Coulomb interaction\cite{peculiar,rpa}.

Recently, a NbSe$_3$ M\"{o}bius strip has been fabricated\cite{tanda}.
It is quite intriguing that a crystalline ribbon forms
this exotic topology. 
Such  the novel experimental finding
motivated us to study the electronic properties of nanographite
ribbons with the \mob boundary condition.
The main purpose of this studty is
to clarify the following questions.
1) How does the \mob boundary
condition affect the peculiar ferri-magnetic states appeared
along the nanographite zigzag edge?
2) Although there have been
some theoretical works on the \mob strip,
these works have been based on the square lattice system.\cite{yakubo,hayashi}
If we make a nano-graphite ribbon the \mob strip form,
what will happen?
In this letter, we tackle on these naive questions 
through the study on the electronic and magnetic properties 
of nanographite M\"{o}bius ribbons.

\begin{figure}[h]
\begin{center}\leavevmode
\includegraphics[width=\linewidth]{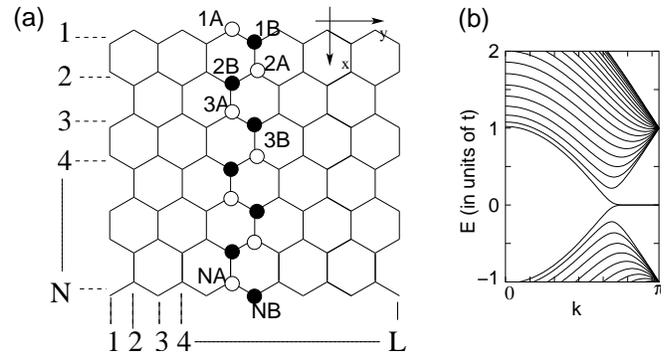}
\caption{(a)The schematic structure
of the zigzag ribbon with finite length $L$.
The length of the ribbon is defined by the number of carbon
slices ($y$-direction), and the width of the ribbon is defined by the
number of carbon atoms which form the each carbon slice.
The 1st carbon slice and the $L${\it th} one are  
connected in terms of the periodic or \mob boundary condition.
(b) The schematic figure of energy band dispersion of
zigzag ribbons, where $N=20$ and $L$ is infinite.
}\label{fig1}\end{center}\end{figure}

Before we discuss the electronic structure of zigzag ribbon with
\mob boundary condition, we shall briefly review the electronic
states of zigzag ribbon.
The nano-graphite ribbon with zigzag edges is obtained 
by making the width $N$ finite and the length $L$ infinite,
in Fig.\ref{fig1}(a).
Throughout this paper, we assume all edge sites are terminated by H-atoms,
also all the $\pi$-electrons have the spherical $s-$orbit for simplicity.
Let us employ the tight-binding model
to study the electronic states of nanographite ribbons.
The tight-binding Hamiltonian is defined by
$H = \sum_{i,j} t_{i,j}c^\dagger_i c_j$,
where $t_{i,j}=-t$ if $i$ and $j$ are nearest neighbors, otherwise 0, 
and $c^\dagger_i(c_i)$ is a creation (annihilation) operator on site $i$.
The ribbon width $N$ is defined by the number of zigzag lines. 
As the graphite lattice is the bipartite, the
A(B)-site on the $n$-$th$ zigzag line is called nA(nB)-site.
Fig. \ref{fig1}(b) is 
the energy band structure of zigzag ribbon for $N=20$,
where the momentum $k$ was introduced along the $y$-direction.
The zigzag ribbons are metallic for all $N$.
One of the remarkable features is the appearance of
partly flat bands at the Fermi level ($E=0$), 
where the electrons
are strongly localized near zigzag edges\cite{peculiar}.

\begin{figure}[h]
\begin{center}\leavevmode
\includegraphics[width=\linewidth]{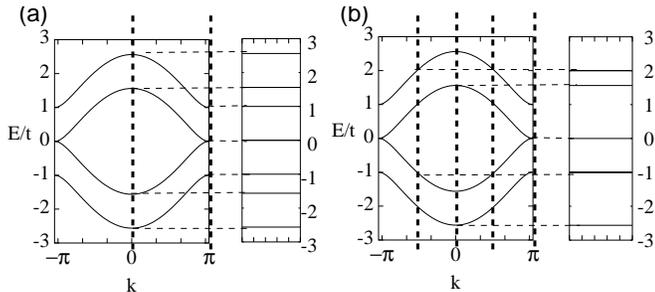}
\caption{(a) The energy band structures of zigzag ribbon of N=4 (left),
and the energy spectrum of nanographite ribbon 
with periodic boundary condition of $N=4$ and $L=4$(right).
(b) Same figures for the \mob boundary condition, where $N=4$ and $L=4$.
The vertical broken lines mean the position of selected wave vector
due to the boundary conditions.
}\label{fig2}\end{center}\end{figure}

The analytic solution of the partly flat band
for a semi-infinite graphite sheet with a zigzag edge 
( 1A and 1B site shall be the edge sites and N is infinite
in Fig.\ref{fig1}(a)) can be expressed as 
$\phi_{nA} = D_k^{n-1}$ and  $\phi_{nB} = 0$,
where $\phi_{nA}$  ($\phi_{nB}$) means the amplitude of the
edge states on nA (nB) site and $D_k = -2\cos(k/2)$.
It is worth noting that the edge state has 
a non-zero amplitude
only on one sublattice, i.e.  meaning the non-bonding character.
Because of the convergence condition of the edge states, 
the wave number $k$ must be $2\pi/3 \le k\le \pi$,
where $|D_k|\le 1$.
In this $k$-region, the edge states make a flat band  
at $E=0$ (Fermi energy).
It should be noted that at $k=\pi$ the edge states are perfectly
localized at the 1A sites, but  at $k=2\pi/3$ the edge states are
completely delocalized.\cite{SSH}
However, when we consider the zigzag ribbons,
two edge states which come from both sides will overlap 
each other and cause the bonding and anti-bonding splitting.
The magnitude of the overlap becomes larger when the wave number 
approaches $2\pi/3$, because the penetration length of the
edge states gets larger there.

Now, we shall consider the electronic states of nanographite ribbon
with the \mob boundary condition.
We consider the nanographite \mob strip  of the geometry with
a rectangle of the length $L$ and the width $N$, requiring its
wave-function $\psi(x,y)$ to satisfy open boundary conditions
in the $x$ direction, and M\"{o}bius boundary conditions in the $y$
direction:
\begin{eqnarray}
\psi(0,y)=\psi(N+1,y)=0, & \\
\psi(x, y+L)=\psi(-x,y). & 
\label{eq:boundary}
\end{eqnarray}
The quantized wave-numbers are $k_x = (\pi/N)n_x$ and 
$k_y=(2\pi/L)([1/2]_{n_x}+n_y)$, where $n_x = 1,2, \cdots$ 
and $n_y = 0,\pm 1,\pm 2, \cdots$. 
The notation $[\alpha]_n$ represents $\alpha$ for $n=$even and $0$ 
for $n=$odd. 
In the cylinder geometry, Eq.(\ref{eq:boundary}) should be replaced by 
$\psi(x,y+L)=\psi(x,y)$, and
gives $k_y=(2\pi/L)n_y$.
Thus, only the $n_x=$even eigenstates are affected by the switch 
from the conventional cylinder (periodic) boundary conditions to the 
M\"{o}bius ones.
These are similar results as the square lattice system.\cite{yakubo}
In Fig.\ref{fig2}, we show the energy levels of nanographite ribbon with
(a) the periodic boundary condition and 
(b) the M\"{o}bius boundary condition.
Since the periodic boundary condition keeps the strip AB-bipartite lattice,
the energy spectrum are symmetric.
However, the zigzag M\"{o}bius strip shows the asymmetric energy spectrum
at $E=0$, because the \mob boundary condition makes the strip
non-AB-bipartite lattice system.
It should be noted that, 
if the width $N$ is even (odd) integer,
the length $L$ has to be even (odd) integer
in order to exclude the polygonal defects such as four-, five-, and
eight-membered rings.

The zigzag ribbon with an infinite length has the strong Fermi instability
by the electron-electron interactions. However, instability 
by the electron-phonon interaction is weak due to the non-bonding
character of the edge states. Here, we employ the mean field Hubbard
model in order 
to study the magnetic structure in the nanographite ribbon with the
\mob boundary condition.
The Hartree-Fock (HF) mean-field hamiltonian is written by
\begin{eqnarray}
H  =  -t
      \sum_{\langle i,j\rangle ,s} 
      c^\dagger_{i,s}c_{j,s}
   + U \sum_{i,s} \left( \langle n_{i,-s}\rangle -1/2\right) n_{i,s}
\end{eqnarray}
where $s$ is spin index.
In terms of the numerical calculation,
we evaluate the electron density at each site, $\langle
n_{i,-s}\rangle$, self-consistently.
Also, the magnetization on each site, $M_i$, is
defined by $M_i = \langle
n_{i,\uparrow}\rangle-\langle n_{i_\downarrow}\rangle$.

\begin{figure}[h]
\begin{center}\leavevmode
\includegraphics[width=0.8\linewidth]{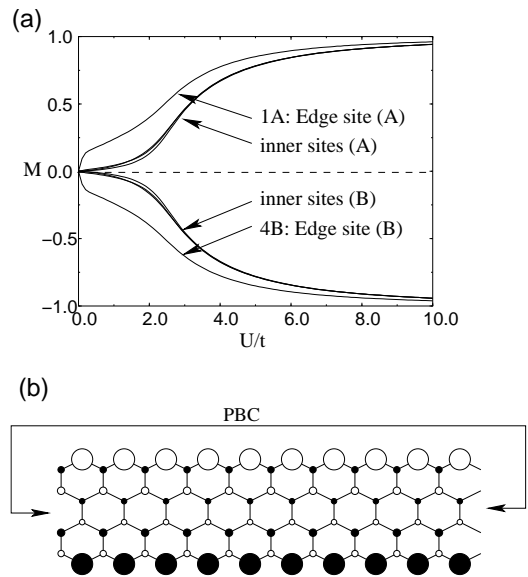}
\caption{(a) The $U/t$ dependence of the magnetization 
in the zigzag ribbon with periodic boundary condition,
where $N=4$ and $L=20$.
(b) The schematic magnetic structure at $U/t=0.1$,
where black and white circles mean the up- and down-spin, respectively.
}\label{fig3}\end{center}\end{figure}
\begin{figure}[h]
\begin{center}\leavevmode
\includegraphics[width=0.8\linewidth]{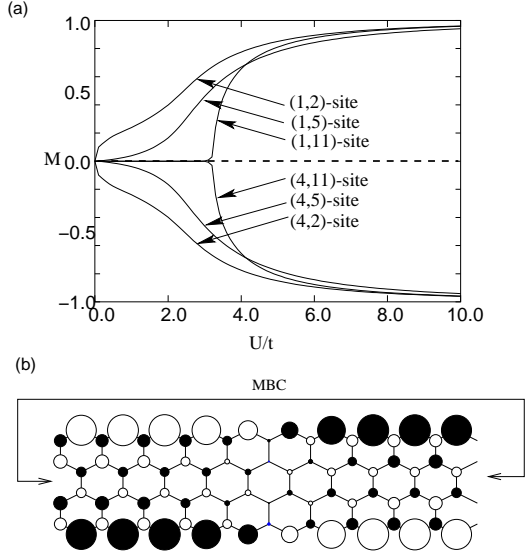}
\caption{(a) The $U/t$ dependence of the magnetization 
in the zigzag ribbon with \mob boundary condition,
where $N=4$ and $L=20$.
(b) The schematic magnetic structure at $U/t=0.1$.
}\label{fig4}\end{center}\end{figure}
\begin{figure}[h]
\begin{center}\leavevmode
\includegraphics[width=0.8\linewidth]{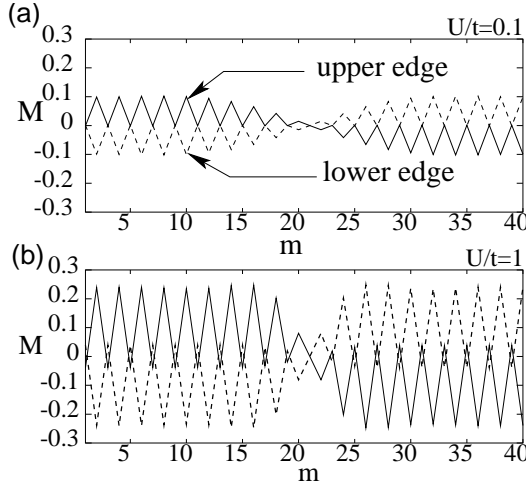}
\caption{
The site dependence of the magnetic moment along the
upper and lower zigzag edges for (a) $U/t=0.1$
and (b) $U/t=1$, where $N=4$ and $L=40$.
}\label{fig5}\end{center}\end{figure}
\begin{figure}[h]
\begin{center}\leavevmode
\includegraphics[width=0.8\linewidth]{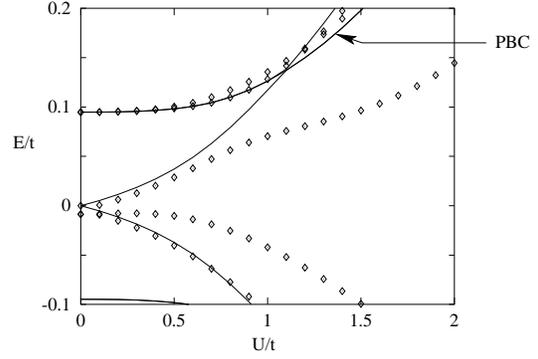}
\caption{The $U/t$ dependence of the energy spectrum.
The solid lines are the results of the graphite ribbon with
periodic boundary condition, and the diamond
represents the results with \mob boundary condition.
}\label{fig6}\end{center}\end{figure}

Previously, we found that this HF Hamiltonian shows
clear differences in the magnetic structure compared with the
graphite sheet\cite{peculiar}.
Since the latter is a zero gap semiconductor, 
where the DOS at the Fermi level is zero, the antiferromagnetic 
HF solution emerges only when $U/t$ is larger than $U_c \sim 2t$,
that is consistent with the fact that the graphite with a very weak 
Coulomb repulsion has no spontaneous magnetism.
The graphite ribbons with zigzag edge, however, display a magnetic
ground state, for any value $U/t > 0$. 
For $U/t \leq 2$, 
magnetic moments appear essentially only at the edge sites
while in the center of the ribbon no magnetism can be found. 
Note that this behavior is consistent with the exact
statement by Lieb for the half-filled Hubbard model\cite{Lieb}.
One of the present authors (K.H.) has discussed the stacking effects 
on the magnetic properties in nanographene planes with zigzag 
edges.\cite{harigaya1,harigaya2}
It has been found that the A-B stackings and the open shell 
electronic structures in each graphene layer are necessary, 
in order to explain the decrease of the magnetic susceptibility
in the course of adsorption of water molecules in activated 
carbon fiber materials.\cite{kawatsu}  In the A-B type stackings, 
the edge sites do not interact with neighboring layers directly,
and the local spin polarizations near edge sites remain againt 
the interlayer hopping interactions.  Therefore, the presence 
of edge sites exhibit an important role, and we can explain 
the novel magnetic properties of nanographites observed recently.

In Fig.\ref{fig3}, we show the $U/t$ dependence of the magnetization 
at each site in zigzag ribbon with the periodic boundary condition.
Of course, the qualitative behavior shares with the
results found in the zigzag ribbons with the infinite length.
Here, the unit of the magnetization is the  Bohr magneton.
For the arbitrary $U/t$, the total magnetization is zero, which is
consistent with the Lieb's theorem, because the
difference between A- and B-sublattice site numbers
is zero. 
However, we find the non-zero solutions of the magnetization 
even if $U/t < 2$, where graphene sheet does not 
have the magnetic solutions.
Since the system has a translational invariance along the
$x$-axis, the magnetization is uniform along the
this direction, as in the infinite zigzag ribbons.
Most remarkable feature in this system is
that the outermost edge sites have largest spin polarization,
which form the ferrimagnetic spin alignment as shown in
Fig.\ref{fig3}(b). 
The variation of the ribbon width and length does not
change the results qualitatively.

Next, we shall discuss the mean-field calculation results for
the \mob strip.
In Fig.\ref{fig4}(a), we show the $U/t$ dependence of the magnetization 
at each site in zigzag ribbon with the \mob boundary condition,
where $N=4$ and $L=20$.
Here we shall change the site index as follows, 
because the system is no longer an AB-bipartite lattice after the
imposition of the \mob boundary condition.
The carbon site on the {\it l}-th zigzag line
and {\it m}-th carbon slice is called ($l$,$m$)-site hereafter.
There are also non-zero solutions of the magnetization when $U/t < 2$
even under the \mob boundary condition.
However, we easily find that the magnetization on 
some sites rapidly decreases around $U/t=4$.
These sites form the magnetic domain wall which crosses the strip as shown
in Fig.\ref{fig4}(b).
The appearance of the magnetic domain wall is somewhat intuitive,
because the \mob boundary condition forces to connect between
same sublattices, resulting in the magnetic
frustration in the ferrimagnetic alignment along a zigzag edge.
Here, we should note that the position of the nodal site is
arbitrary, because the \mob strip is originally translational
invariant under $U/t=0$.
%
%
%
In Fig.\ref{fig5}, we show the site dependence of the magnetic moment 
along the two zigzag edges for the $U/t=0.1$ and $1$. We clearly observe that
the width of the domain wall depends on the 
magnitude of the Coulomb interactions.
Increasing $U/t$, the width of the domain wall is gradually narrowing.
Here, we shall comment on the total magnetization 
of the nanographite \mob strip system. We found that the total
magnetization of the \mob strip with even (odd) $N$ is zero (non-zero).
In the original rectangle system without \mob boundary condition,
the difference between A- and B-sublattice site numbers is zero 
(non-zero) for the even (odd) $N$ system.
Although the \mob boundary condition makes the system a non-AB-bipartite, 
this fact reminds us that the argument of the Lieb's theorem
is still valid under the \mob boundary condition.

Finally, we show the $U/t$ dependence of the energy spectrum
for $N=4$ and $L=20$.
In Fig.\ref{fig6},
the solid lines indicate the variation of the energy spectrum
under the periodic boundary condition,
so that the Coulomb interaction produces the energy gap.
However, under the \mob boundary condition,
new energy states appear inside the Hubbard gap.
This is quite similar to the mid-gap states appeared in the 
doped polyacetylene systems due to the bond alternation, i.e. the 
soliton level. \cite{heeger}

In conclusion,
we have studied the electronic 
and magnetic properties of the nanographite 
ribbon with zigzag edges under the periodic or M\"{o}bius boundary
conditions. The zigzag nano-graphite ribbons possess edge localized
states at the Fermi level which cause a ferrimagnetic spin polarization
localized at the edge sites even in the very weak Coulomb interaction. 
The imposition of the \mob boundary condition makes the system
an non-AB-bipartite lattice, and depress the spin 
polarization at the boundary of the magnetic domain, 
resulting in the formation of a magnetic domain wall.
The width of the magnetic domain depends on the Coulomb interaction
and narrows with increasing $U/t$. 
However, there remain problems which have not been
treated here.
In this paper, we have assumed that each $\pi$-electron has
the spherical {\it s}-orbital, however in reality $\pi$-electron
has the {\it p}-orbital. The electronic states of
the nanographite \mob strip with the {\it p}-orbitals will be presented
elsewhere. The effect on the electronic states
by Bohm-Aharanov magnetic flux, passing through
the \mob strip, is also intriguing and
will be presented elsewhere.

One of the authors (K.W.) is grateful to S. Okada for his helpful discussion.
He is also grateful for support by Grant-in-Aid for Scientific
Research from Ministry of Education, Science and Culture, Japan.
and the financial support from the Foundation Advanced Technology 
institute and from the Kinki-chihou Hatsumei Center.
The other author (K.H.) is grateful to T. Enoki, K. Takai, N. Kawatsu, 
and Y. Kobayashi for useful discussion on experimental aspects of 
nanographite materials.  He also acknowledges the financial support by the
Synthetic Nano-Function Materials Project from NEDO.

\end{document}